\documentclass[useAMS]{gGAF2e}
\usepackage{graphicx}
\usepackage{amsmath}
\newcommand{\pa}{\partial}
\newcommand{\ob}{\overline}
\newcommand{\beq}{B_{\rm eq}}
\newcommand{\erf}{\textrm{erf}}
\newcommand{\Rs}{R_{\odot}}
\newcommand{\Rm}{R_{\rm m}}
\begin{document}
\doi{}
 \issn{} \issnp{} \jvol{00} \jnum{00} \jyear{2010} 
\markboth{Chatterjee, Brandenburg and Guerrero}{Dynamical quenching in $\alpha\Omega$ dynamos}

\title{Can catastrophic quenching be alleviated by separating shear and $\alpha$ effect?}

\author{Piyali Chatterjee, Axel Brandenburg and Gustavo Guerrero}

\maketitle

\begin{abstract}
The small-scale magnetic helicity produced as a by-product of the
large-scale dynamo is believed to play a major role in
dynamo saturation. In a mean-field model the generation of small-scale
magnetic helicity can be modelled by using the dynamical quenching
formalism. Catastrophic quenching refers to a decrease of the saturation
field strength with increasing Reynolds number. It has been
suggested that catastrophic quenching 
only affects the region of non-zero helical turbulence
(i.e.\ where the kinematic $\alpha$ operates) and that it is possible to
alleviate catastrophic quenching by separating the
region of strong shear from the $\alpha$ layer. We perform a
systematic study of a simple axisymmetric two-layer $\alpha\Omega$ dynamo
in a spherical shell for Reynolds numbers in the range
$1 \leq \Rm \le 10^5$. In the framework of dynamical quenching we show that
this may not be the case, suggesting that magnetic helicity fluxes would be necessary.
\end{abstract}


\section{Introduction}
It is widely believed that the solar magnetic cycle is
driven by an $\alpha\Omega$ dynamo. The region of strong radial
shear called the tachocline \citep{SZ92}
at the bottom of the solar convection zone (SCZ) is believed to
be the place where strong toroidal field is formed due to 
stretching of the weaker but diffuse poloidal field. It has been 
inferred from helioseismology that the tachocline is confined
within a thin layer in the overshoot region which lies below the SCZ.
This led \cite{Par93} to propose the idea of an interface dynamo,
where the shear is confined to a region with a greatly reduced
turbulent diffusion, which also is the region of production of 
strong toroidal field. The helical turbulence generated due to
convection and rotation in the layer above provides the
turbulent $\alpha$ and a large turbulent diffusivity $\eta_{\rm t}$.
The dynamo cycle being thus completed, the interface dynamo
operates as a surface wave propagating along the boundary between
strong shear and convection, which is also a region with a strong
gradient in the turbulent diffusivity.

In order to mimic the process of dynamo saturation, traditionally
authors have used an algebraic quenching function like 
$\alpha_0/(1+\overline{\bm{B}}^2/B_{\rm eq}^2)$ or even
$\alpha_0/(1+R_{\rm m} \overline{\bm{B}}^2/B_{\rm eq}^2)$, where
$\alpha_0$ is the unquenched value of $\alpha$, $\Rm$ is
the magnetic Reynolds number, $\overline{\bm{B}}$ is the mean magnetic field and
$B_{\rm eq}$ is the equipartition magnetic field.
However, 
it appears that conservation of magnetic helicity plays an important
role in the process of saturation. At large $\Rm$, the
magnetic helicity ($\int\bm A\cdot\bm B\, dV$) is fairly well
conserved by the dynamo producing equal amounts of helicity in small
and large scales,
respectively. If the small-scale magnetic helicity is not able to escape
out of the system, then the turbulent  $\alpha$ effect is markedly
reduced \citep{PFL}. 
This leads to the magnetic energy of the dynamo to be quenched
such that the saturation value varies as $\Rm^{-1}$. 
Since astrophysical objects have large $\Rm$ ($\Rm\sim10^9$ for the Sun), this strong
dependence of the saturation field strength $B_{\rm sat}$ on $\Rm$ 
is referred to as
{\em catastrophic quenching}.

Even though the helicity 
constraint in direct numerical simulations (DNS) 
of dynamos with strong shear has been clearly identified, the results   
can be matched with mean-field models having a weaker algebraic
quenching of $\alpha$ and turbulent diffusivity than $\alpha^2$ dynamos \citep{BBS01}.
However, the empirically determined coefficients would depend on circumstances
and are therefore not universal.
Such a model would not obey magnetic helicity evolution and is therefore
untenable on theoretical grounds.
The interface dynamo 
model has been invoked several times as a way of 
getting around the persistent problem of  catastrophic 
quenching \citep[see][for a review]{Cha05}. The belief is that
the quenching function remains close to unity in the region of finite
$\alpha$ since the toroidal field is expected to be weak there.
However, to our knowledge, this has never been verified in a
consistent manner for a range of magnetic Reynolds numbers.
In this paper we perform a series of calculations with mean-field 
$\alpha\Omega$ models, in spherical geometry, considering both
algebraic and dynamical quenching formulations, for magnetic Reynolds
numbers in the range $1\le \Rm \le 2\times10^5$. 
An important feature of these models is that the region of strong 
narrow shear is separated from the region of helical turbulence
as proposed in the Parker's interface model. 

In \S2 we discuss the features of the $\alpha\Omega$ model used, and
the formulation of  dynamical $\alpha$ quenching. The results are
highlighted in \S3 and  conclusions are drawn in \S4.  A part of
the calculations presented in this paper will be discussed in
a more detailed paper \citep{CGB10}.
In this paper we focus specifically
on the reality of the catastrophic quenching in dynamos
with $\alpha$ and $\Omega$ effects operating in two widely separated
layers.

\section{Nonlinear $\alpha\Omega$ Dynamo}

\subsection{The underlying mean-field model}

Our dynamo model consists of the induction equations for the toroidal
component of the mean
poloidal field potential, $A_{\phi}(r, \theta)$,
and the mean toroidal magnetic field,
$B_{\phi}(r, \theta)$, written in spherical geometry
under the assumption of axisymmetry ($\pa/\pa \phi = 0$).
In some of the cases an additional evolution equation will be solved
for the $\alpha$ effect (described in the next section).
We have used a modified version of the publicly available solar dynamo code {\em
  Surya}\footnote{The code {\em Surya} and its manual can be obtained
  by writing an email to {arnab@physics.iisc.ernet.in}} described in
\cite{CNC04} to perform these calculations. 

In this paper we have used a smoothed step
profile for 
$\eta$ given by 
\begin{equation}
\label{eq:eta}
\eta(r)=\eta_r+\frac{1}{2} \eta_{\rm
  t}\left[1+\erf\left(\frac{r-r_{e}}{d_e}\right)\right], 
\end{equation}
where $r_{e}=0.73\Rs$ and $d_e=0.025\Rs$. We define 
the magnetic Reynolds number as $\Rm=\eta_{\rm t}/\eta_r$. 
In order to facilitate comparison with earlier work in Cartesian
geometry \citep[see, e.g.,][]{BCC09}, it is convenient to define
an effective minimal wavenumber, $k_1$.
Somewhat arbitrarily, we use 
$k_1=2/\Rs$, which corresponds to a harmonic wave with 2 nodes
spanning the full latitudinal extent between both poles.
We also define the wavenumber of the energy-carrying eddies, $k_{\rm f}$,
corresponding to the inverse pressure scale height near the base of
the convection zone. For all our calculations we have
taken $k_{\rm f}=7k_1$.
Using the estimate $\eta_{\rm t}=u_{\rm rms}/3k_{\rm f}$ we
can express the equipartition field strength with respect to the
turbulent kinetic energy as
\begin{displaymath}
\beq = (4\pi\rho)^{1/2}u_{\rm rms}= (4\pi\rho)^{1/2}\, 3\eta_{\rm t} k_{\rm f}.
\end{displaymath}
For algebraic quenching there is no magnetic $\alpha$ effect and
we just have the kinetic $\alpha$ effect, $\alpha_{\rm K}$,
which we assume to be of the form
\begin{equation}
\label{eq:alphak}
\alpha_{\rm K}(r)=\left. \frac{1}{2} \alpha_0 \cos\theta
\left[1+\erf\left(\frac{r-r_a}{d_a}\right)\right] \right/
\left({1+g_{\alpha}\ob{B}^2/\beq^2}\right),
\end{equation}
where the value of $\alpha_0$ may be computed using the
first order smoothing approximation (FOSA)
as being equal to $\epsilon_{\rm f}\eta_{\rm t} k_{\rm f}$ \citep{BB02}.
Here, the prefactor $\epsilon_{\rm f}$ is usually of order 0.1
or less since $({\bm{u}\cdot\bm{\omega}})_{\rm rms} < u_{\rm
  rms}\omega_{\rm rms}$. The case $\epsilon_{\rm f}=1$ means the flow
is maximally helical. The term 
$g_{\alpha}$ is a non-dimensional coefficient equal to 1 or 
$\Rm$ depending on the assumed form of algebraic quenching in
the models.  Even though the helical turbulence pervades almost the
entire  convection zone, we take $r_a = 0.77\Rs$ and $d_a=0.015\Rs$ so
that we can have a large separation between the shear layer and
the layer where turbulence is important.
Consequently we consider a differential rotation profile like
that  in the high latitude tachocline of the Sun given by,
\begin{equation}
\label{eq:omega}
\Omega(r)=-\frac{1}{2}\Omega_0\left[1+\erf\left(\frac{r-r_w}{d_w}\right)\right],
\end{equation} 
where $\Omega_0=14$nHz, $r_w=0.68\Rs$ and $d_w=0.015\Rs$. The radial 
profiles of $\eta_{\rm t}$, $\alpha$ and $\pa \Omega/\pa r$ are
plotted as a function of fractional radius $r/\Rs$ in
Fig.~\ref{fig:profiles}.  The region of strong radial shear is
thus separated from the region of helical turbulence and the
diffusivity has a strong gradient at a radius lying between 
the two layers.
It may be noted that,
in order to have supercritical dynamo action 
in a model with $\eta$ having the same radial profile as $\alpha$,
we must set $\epsilon_{\rm f} \gg 1$. If the strong gradient of $\eta$
lies between the two source regions, then we can work with
$\epsilon_{\rm f} \leq 1$. Also the time period $T_{\rm cyl}$
of the oscillatory dynamo remains a reasonably small fraction of the
turbulent diffusion time $t_{\rm diff}$. We can justify the profiles of $\eta$ and 
$\alpha$ on the grounds of having dynamo action for a reasonable range of parameters even while 
avoiding any significant overlap between the two source regions.
The formulation of the equation for the evolution of the $\alpha$ effect
for dynamical quenching is described in \S2.2.

\begin{figure}
  \label{fig:profiles}
  \centering
  \includegraphics[width=0.6\textwidth]{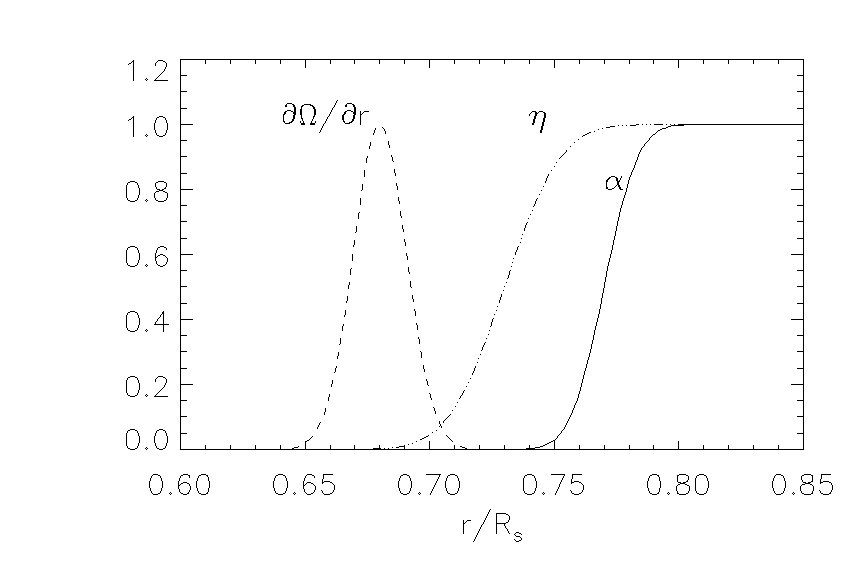}
  \caption{\label{fig:profiles} Profiles of negative radial shear
    $\pa \Omega/ \pa r$ (dashed), $\alpha$ (solid) and 
    $\eta$ (dashed-dotted) as a function of fractional solar radius.}%
\end{figure}
\subsection{Dynamical $\alpha$ quenching}
It was first shown by \cite{PFL} that the turbulent
$\alpha$ effect is modified due to the generation of small-scale
magnetic helicity such that the total $\alpha$ effect is
given by,

\begin{equation}
  \label{eq:alpha}
  \alpha=\alpha_{\rm K}+\alpha_{\rm M}=-\frac{\tau}{3}\left(\overline{\bm\omega\cdot\bm{u}} -
  \rho^{-1}\overline{\bm{j\cdot b}}\right), 
\end{equation}
where $\bm\omega$, $\bm{u}$, $\bm{j}$, $\bm{b}$ denote the fluctuating
components of vorticity, velocity, current density, and magnetic field in
the plasma, respectively.

For this type of quenching 
we use the same radial and latitudinal profiles of $\alpha_K$ as given in
Eq.~(\ref{eq:alphak}), but without the algebraic quenching factor in
the denominator, i.e.\ we put $g_\alpha=0$.
The second term in Eq.~(\ref{eq:alpha}) is sometimes referred to as
the magnetic $\alpha$ effect or $\alpha_{\rm M}$. 
It is  possible to write an equation for the evolution of $\alpha_{\rm M}$
from the equation for the evolution of the small-scale  magnetic
helicity density $h_{\rm f}=\overline{\bm{a}\cdot \bm{b}}$ using the
relation \citep{BCC09},
\begin{equation}
\label{eq:rel}
\alpha_{\rm M} = \frac{\eta_{\rm t} k_{\rm f}^2}{B_{\rm eq}^2}h_{\rm f}.
\end{equation}
The equation for $\overline{\bm{a\cdot b}}$ is in principle
gauge-dependent.
However, under the assumption of scale separation, i.e.\ when the
correlation length of the turbulence is small compared to the system size,
one can define a magnetic helicity density of small-scale fields in
a gauge-independent manner as the density of linkages \cite{SB06}.
Using Eq.~(\ref{eq:rel}), this leads to an evolution equation for $\alpha_{\rm M}$,
\begin{equation}
\label{eq:alphaeq}
\frac{\pa \alpha_{\rm M}}{\pa t} = -2\eta_{\rm t} k_{\rm
  f}^2\left(\frac{\overline{\bm{\mathcal{E}}}\cdot\overline{\bm{B}}}{B_{\rm
    eq}^2}+\frac{\alpha_{\rm M}}{R_{\rm m}}\right)-\nabla\cdot
{\bm F}_{\alpha}, 
\end{equation}
where $\bm{\mathcal{\overline{E}}}$ and $\overline{\bm B}$ are the 
mean  electromotive force and the mean magnetic field. 
The flux ${\bm F}_{\alpha}$ 
consists of individual contributions, e.g., 
advection due to the mean flow, the Vishniac--Cho flux
\citep{VC01,SB04}, diffusive fluxes, 
triple correlation terms, etc. 
The effect of some of these individual fluxes in a spherical geometry
and with both radial and latitudinal shear will be investigated in a
paper under preparation \citep{Gue10}, but for all the simulations 
presented in this paper we have put ${\bm F}_{\alpha}=0$.

We solve the equations for  $A_{\phi}(r,
\theta)$, $B_{\phi}(r, \theta)$ and $\alpha_{\rm M}$ 
in a domain confined by $0 \le \theta \le \pi$ and $0.55\Rs\le r \le 
\Rs$. The boundary conditions for $A_{\phi}$ are given by a 
potential field condition at the surface \citep{DC94}
and $A_{\phi}=0$ at the poles. At the bottom we use the perfect
conductor boundary condition $\pa (r B_{\theta})/\pa r =
\pa(rB_{\phi})/\pa r=0$.  Also $B_{\phi}=0$ on all other boundaries. 
We have checked that the results are not very sensitive to 
different boundary conditions 
at the bottom boundary mainly because the boundary is far removed from
the dynamo region. 
Since ${\bm F}_{\alpha}=0$, no derivatives of $\alpha_{\rm M}$ need to be evaluated,
so no boundary conditions need to be specified for $\alpha_{\rm M}$ and its
evolution equation is just an initial value problem.
We start with an initial dipolar solution where 
$B_{\phi}$ is antisymmetric about the equator.

\section{Results}
To study the $\Rm$ dependence of $B_{\rm sat}$ in our model we keep all
the dynamo parameters the same for all the runs except $\eta_r$
which we change from $2\times10^5$ cm$^2$ s$^{-1}$ to $2\times10^{10}$ cm$^2$ s$^{-1}$
while keeping $\eta_{\rm t}$
fixed at $4\times10^{10}\,{\rm cm}^2\,{\rm s}^{-1}$. It may also be noted that 
the time period of the dynamo models ($T_{\rm cyl}$) is fairly 
independent of the magnetic Reynolds number. 

\begin{figure}
   \centering{\includegraphics[width=0.5\textwidth]{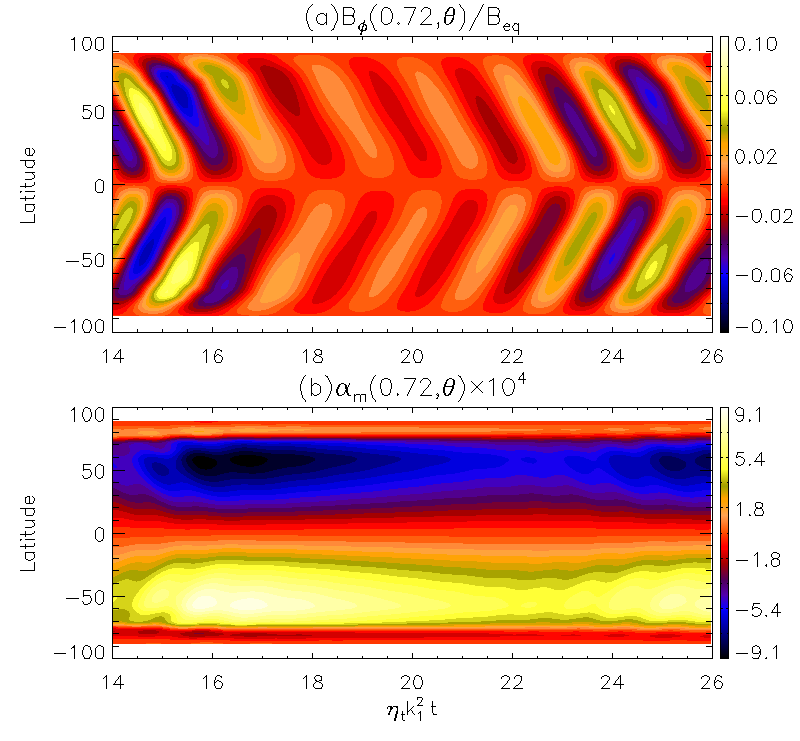}}
      \caption{(a) $B_{\phi}(0.72\Rs,\theta)$ and (b) $\alpha_{\rm m}(0.72\Rs,\theta)$ as a function of diffusion time $\eta_{\rm t} k_1^2t$ for $R_{\rm m}=2\times10^3$
              }
         \label{fig:butter2}
   \end{figure}

\begin{figure}
      \centering{\includegraphics[width=0.8\textwidth]{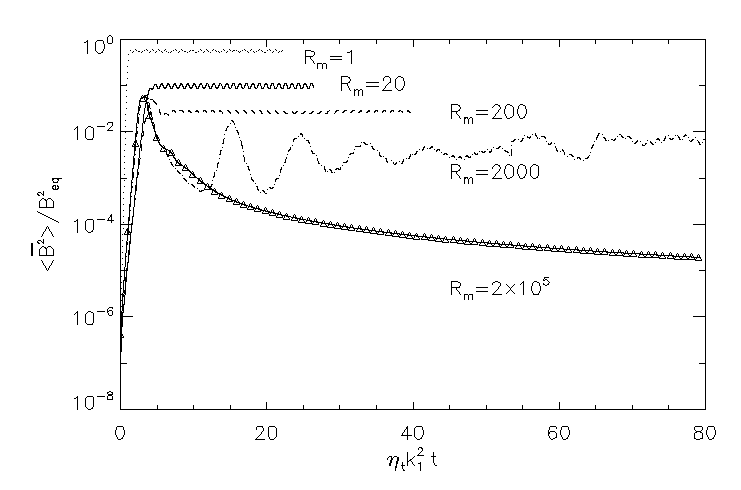}}
      \caption{Volume averaged magnetic energy in the domain scaled with the equipartition energy for $R_{\rm m}=1$ (dotted line), $R_{\rm m}=20$ (solid), $R_{\rm m}=200$ (dashed), $R_{\rm m}=2\times10^3$ and $R_{\rm m}=2\times10^5$ 
(triangles) with dynamical quenching. Adapted from \cite{CGB10}.
              }
         \label{fig:energy3}
   \end{figure}

To be able to correctly compare the dynamo models for different $R_{\rm m}$, 
we have calculated the critical value of $\alpha_0$,
denoted by $\alpha_{\rm c}$ for each model. 
In the following we present results for $\alpha_0=2\alpha_{\rm c}$.
We show in Fig.~\ref{fig:butter2} the butterfly diagrams for $B_{\phi}$
and $\alpha_{\rm M}$ at a depth of 0.72$\Rs$.
It may be concluded from the butterfly diagram for $\alpha_{\rm M}$
in Fig.~\ref{fig:butter2}b
that the small-scale current helicity, and hence $\alpha_{\rm M}$,
is predominantly negative
(positive) in the Northern (Southern) hemisphere. Let us denote 
the exponential decay time for $\alpha_{\rm M}$ by $t_{\alpha}$. So, 
$t_{\alpha} = \Rm/\eta_{\rm t}k_{\rm f}^2 = 4.55\times 10^{-3}\Rm t_{\rm diff}$.
For $R_{\rm m}=2\times10^3$, the decay time $t_{\alpha} \gg T_{\rm cyl}$ 
and so the system of equations is overdamped, as can be seen from the 
butterfly diagrams in Fig.~\ref{fig:butter2}a as well as from 
saturation curve (dashed dotted line) in Fig.~\ref{fig:energy3}.
Note that there are amplitude modulations of the magnetic field before it 
settles to a final saturation value. The nature of the saturation curves of the magnetic energy is thus strongly
governed by the ratio of $t_{\alpha}$ and $T_{\rm cyl}$.

\begin{figure}
      \centering{\includegraphics[width=0.6\textwidth]{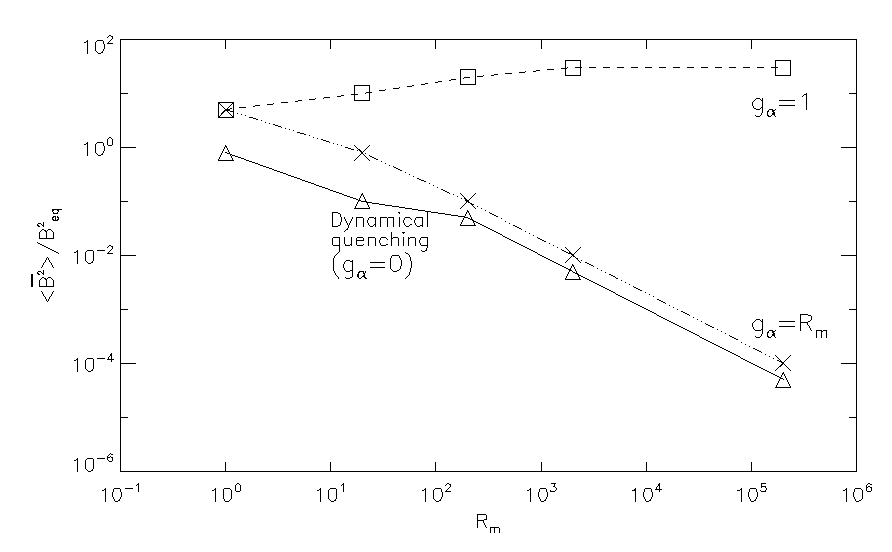}}
      \caption{Volume averaged magnetic energy scaled with the
      equipartition energy  in the saturation phase as a function of
      $R_{\rm m}$ for dynamical $\alpha$ quenching (triangles +solid)
      and algebraic quenching with $g_{\alpha}=1$ (squares + dashed)
      and with $g_{\alpha}=R_{\rm m}$ (cross + dashed-dotted).
      Adapted from \cite{CGB10}.  
              }
         \label{fig:satrm}
   \end{figure}

The results of our calculations for different Reynolds numbers are 
plotted in Fig~\ref{fig:energy3}. The slopes in the kinematic phase are 
similar for all $R_{\rm m}$ within the error in the numerical 
determination of the critical $\alpha_c$. The strong $R_{\rm m}$ dependence,
which is reminiscent of catastrophic quenching in large $\Rm$
dynamos, can be easily discerned from the same figure,
but is more clear in Fig.~\ref{fig:satrm} where we see that 
the saturation energy decreases monotonically as a function of
magnetic Reynolds number.
For $R_{\rm m} =2\times10^5$, 
the code has to be run for 500 $t_{\rm diff}$ before the dynamo fields
may start  
becoming `strong' again for the case with $\alpha_0=2\alpha_c$. 
Due to long computational times involved in this exercise we have not 
continued the calculation beyond 60 $t_{\rm diff}$.
Hence the determination of saturation magnetic energy may be
inaccurate for $R_{\rm m}=2\times10^5$. 
In Fig.~\ref{fig:satrm} we compare the case with dynamical quenching
against the cases 
with a simple algebraic quenching of the 
form given in Eq.~(\ref{eq:alphak}) with $g_{\alpha}=1$ and
$g_{\alpha}=R_{\rm m}$. 
We notice that 
for $g_{\alpha}=R_{\rm m}$, the algebraically and dynamically quenched 
$\alpha$ effects seem to give similar dependences on $R_{\rm m}$.

 When the code is run longer, 
we start seeing changes in the parity after $t > 40\, t_{\rm diff}$
for the dynamically quenched system in contrast to the ´strong' 
algebraic quenching case, where the parity remains dipolar.
However the magnetic energy and the dynamo period $T_{\rm cyl}$ remain
fairly constant even while the system fluctuates between symmetric 
and anti-symmetric parity at an irregular time interval. 

\begin{figure}
\centering{\includegraphics[width=0.5\textwidth]{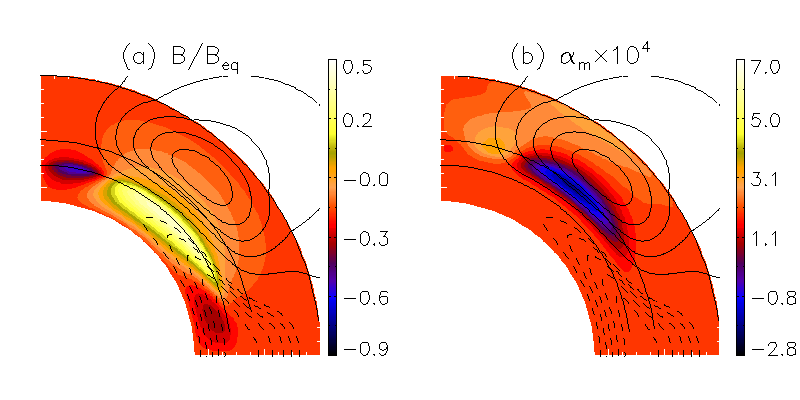}
    \includegraphics[width=0.5\textwidth]{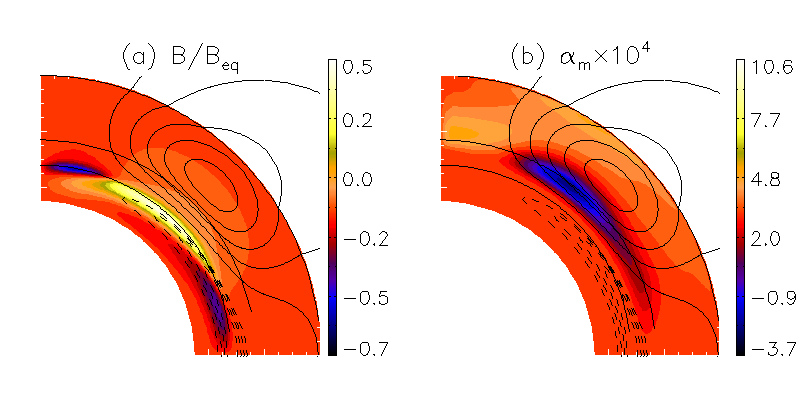}
     \includegraphics[width=0.5\textwidth]{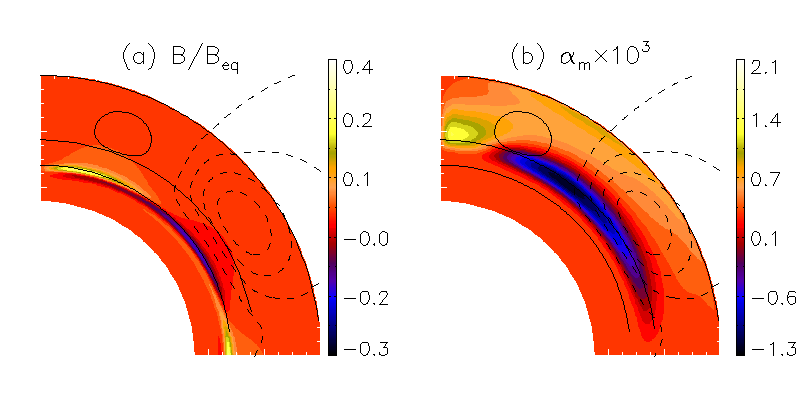}}
%      \caption{Meridional snapshots of (a) $B/B_{\rm eq}$ and (b) $\alpha_{\rm m}$ for cases with different $R_{\rm m}$ starting from 20 (upper panel), 200 (middle panel) and 2$\times10^3$ (lower panel). Two concentric circles have been drawn at 0.68$\Rs$ and 0.77$\Rs$ to denote the radial positions of the shear layer and the $\alpha$ effect.
%AB: added "(color/grey-scale coded)". Also replaced B -> B_phi for clarity,
%AB: because we do this elsewhere in the text, even though the
%AB: figure just says B, but that's ok.
      \caption{Meridional snapshots of (a) $B_\phi/B_{\rm eq}$ and
      (b) $\alpha_{\rm m}$ (color/grey-scale coded) for cases with
      different $R_{\rm m}$ starting from 20 (upper panel),
      200 (middle panel) and 2$\times10^3$ (lower panel).
      Two concentric circles have been drawn at 0.68$\Rs$ and
      0.77$\Rs$ to denote the radial positions of the shear layer and
      the $\alpha$ effect.
      %The solid (dashed) contours denote the positive (negative) poloidal field potential.
%AB: I think this is more accurate
      Solid and dashed lines denote poloidal field lines, corresponding to
      contours of positive (negative) $r\sin\theta A_\phi$.
              }
         \label{fig:snap}
   \end{figure}

In Fig.~\ref{fig:snap} we show the meridional snapshots in the Northern 
hemisphere of the toroidal component of the magnetic field and $\alpha_{\rm M}$.
It may be noted that the regions strongest in $B_{\phi}$ become
progressively confined in the narrow shear layer with increasing
$R_{\rm m}$  while $\alpha_{\rm M}$
becomes stronger leading to decrease in $B_{\rm sat}$.
Even though $\alpha_{\rm M}$ is predominantly negative in the Northern
hemisphere there is a region of positive small-scale helicity generated
just below the region where $\alpha_{\rm K}$ is finite so that contribution
from the term $\alpha B_{\phi}^2$ in the source 
$\bm{\mathcal{\overline{E}}}\cdot \ob{\bm{B}}$
is small.
This effect is similar to the one reported in \cite{BCC09}.
We also have not observed any 
evidence of chaotic behaviour in the range of magnetic Reynolds 
number $20 \le R_{\rm m} \le 2\times 10^5$ for twice supercritical 
$\alpha$ ($\alpha\leq 2\alpha_{\rm c}$), in agreement with \cite{Covas}

\section{Conclusions}
The calculations done in this paper indicate that it is not 
possible to escape catastrophic quenching due to accumulation of small-scale helicity in the domain by merely separating the regions of 
shear and $\alpha$ effect. 
The saturation value of magnetic energy decreases as $\sim R_{\rm m}^{-1}$ for 
both dynamical quenching and the `strong' ($\sim R_{\rm m}$-dependent)
form of algebraic quenching for 
the simple two-layer model. However, additionally we observe parity 
fluctuations for cases with dynamical quenching.
It does not seem to us that there exists  
any chaotic behaviour in the time series of magnetic energy since the dynamo
period and the saturation energy remains fairly constant. 
It may be possible that solar wind, coronal mass ejections,
and Vishniac and Cho fluxes help in throwing out the small  
scale helicity from the Sun and thus alleviate catastrophic quenching. 
In this study we have not found any difference between the nature
of the saturation curves for an $\alpha\Omega$ dynamo and an $\alpha^2$
dynamo using the form of dynamical
quenching given by Eq.~(\ref{eq:alphaeq}).
However, it is clear that the algebraic quenching formula must fail
if we were to allow for magnetic helicity fluxes that would, under
suitable circumstances, alleviate catastrophic quenching.

We have been cautious about using dynamical quenching equation for 
dynamo numbers not very large compared to the critical dynamo numbers.
We would expect that the magnetic field
should affect all the turbulence coefficients including both
$\alpha$ and $\eta$. However for this analysis we have not included 
any quenching of $\eta_{\rm t}$.
This may be justified since such an effect could be mimicked by our 
simple two-layer model with a lower $\eta_{\rm t}$ in the region of 
production of strong toroidal fields and a higher $\eta_{\rm t}$ in the 
region of weaker poloidal fields.
The effect of dynamical quenching on more realistic solar dynamo models
having meridional circulation, Babcock-Leighton $\alpha$ effect,
and diffusive helicity fluxes have also been studied \citep{CGB10}.

Unfortunately the direct numerical simulations have not yet reached the
modest Reynolds numbers used in this paper ($\sim 10^4$) which are still much lower 
than the astrophysical dynamos. 
If $B_{\rm sat}^2$ really has an inverse dependence on $\Rm$, then the solar 
dynamo should not be operating like it does! We may conclude that either 
we must have helicity fluxes out of the system or cross-equatorial
diffusive fluxes inside the domain or that 
the $\alpha_{\rm M}$ equation to be used beside the mean-field induction 
equations must be suitably modified for $\alpha\Omega$ dynamos. 
To verify if the equation for dynamical 
quenching works in the same way as in $\alpha^2$ dynamos, we need to 
perform systematic comparisons between DNS with shear and convection 
and mean-field modelling for $\alpha\Omega$ dynamos.

\section*{Acknowledgements}
P.C. and A.B. would like to thank
the organisers of the `Natural Dynamos' meeting in High Tatras, 
where this work was initiated. 
We acknowledge the allocation of computing resources provided by the
Swedish National Allocations Committee at the Center for
Parallel Computers at the Royal Institute of Technology in
Stockholm and the National Supercomputer Centers in Link\"oping
as well as the Norwegian National Allocations Committee at the
Bergen Center for Computational Science. 
This work was supported in part by
the European Research Council under the AstroDyn Research Project No.\ 227952
and the Swedish Research Council Grant No.\ 621-2007-4064.

\newcommand{\yana}[5]{, ``#5,'' {\em Astron.\ Astrophys.\ }{\bf #2}, #3-#4 (#1).}
\newcommand{\yanaS}[5]{, ``#5'' {\em Astron.\ Astrophys.\ }{\bf #2}, #3-#4 (#1).}
\newcommand{\yapj}[5]{, ``#5,'' {\em Astrophys.\ J.\ }{\bf #2}, #3-#4 (#1).}
\newcommand{\yjfm}[5]{, ``#5,'' {\em J.\ Fluid Mech.\ }{\bf #2}, #3-#4 (#1).}
\newcommand{\ymn}[5]{, ``#5,'' {\em Monthly Notices Roy.\ Astron.\ Soc.\ }{\bf #2}, #3-#4 (#1).}
\newcommand{\ysph}[5]{, ``#5,'' {\em Solar Phys.\ }{\bf #2}, #3-#4 (#1).}
\newcommand{\ypre}[5]{, ``#5,'' {\em Phys.\ Rev.\ E }{\bf #2}, #3-#4 (#1).}
\newcommand{\yprlN}[5]{, ``#4,'' {\em Phys.\ Rev.\ Lett. }{\bf #2}, #3 (#1).}
\newcommand{\yjourN}[5]{, ``#5,'' {\em #2} {\bf #3}, #4 (#1).}

\end{document}